%Paper: hep-th/9511030
%From: Edward Witten <witten@IAS.EDU>
%Date: Sat, 04 Nov 1995 11:29:01 EST

\input harvmac
\newcount\figno
\figno=0
\def\fig#1#2#3{
\par\begingroup\parindent=0pt\leftskip=1cm\rightskip=1cm\parindent=0pt
\baselineskip=11pt
\global\advance\figno by 1
\midinsert
\epsfxsize=#3
\centerline{\epsfbox{#2}}
\vskip 12pt
{\bf Fig. \the\figno:} #1\par
\endinsert\endgroup\par
}
\def\figlabel#1{\xdef#1{\the\figno}}
\def\encadremath#1{\vbox{\hrule\hbox{\vrule\kern8pt\vbox{\kern8pt
\hbox{$\displaystyle #1$}\kern8pt}
\kern8pt\vrule}\hrule}}

\overfullrule=0pt

%macros
%

\def\bar{\overline}
\def\Z{{\bf Z}}
\def\T{{\bf T}}
\def\S{{\bf S}}
\def\R{{\bf R}}

\font\zfont = cmss10 %scaled \magstep1

\def\bigone{\hbox{1\kern -.23em {\rm l}}}
\def\ZZ{\hbox{\zfont Z\kern-.4emZ}}

\Title{hep-th/9511030, IASSNS-HEP-95-87}
{\vbox{\centerline{SMALL INSTANTONS IN STRING THEORY}}}
\smallskip
\centerline{Edward Witten\foot{Research supported in part
by NSF  Grant PHY92-45317.}}
\smallskip
\centerline{\it School of Natural Sciences, Institute for Advanced Study}
\centerline{\it Olden Lane, Princeton, NJ 08540, USA}\bigskip

\medskip

\noindent
%write abstract here
A long-standing puzzle about the heterotic string has been
what happens when an instanton shrinks to zero size.
It is argued here that the answer at the quantum level
is that an extra $SU(2)$ gauge
symmetry appears that is supported in the core of the instanton.
Thus in particular  the quantum  heterotic string has
vacua with higher rank than is possible in conformal field theory.
When $k$ instantons collapse at the same point, the enhanced
gauge symmetry is $Sp(k)$.
These results, which can be tested by comparison to Dirichlet
five-branes of Type I superstrings and to the ADHM construction
of instantons, give  the first example for the heterotic
string of a non-perturbative phenomenon that cannot be turned off
by making the coupling smaller.  They have applications
to several interesting puzzles about string duality.

\Date{October, 1995}
%text of paper

\newsec{Introduction}

One of the surprises in studies of string dynamics in the last
year has been that there are non-perturbative effects which
-- though beyond reach of conformal field theory -- occur
no matter how small the string coupling constant may be.
For instance, the Type IIA superstring compactified on K3 gets
an extended gauge symmetry when a two-sphere collapses to zero
size \ref\witten{E. Witten, ``String Theory Dynamics In Various
Dimensions,'' Nucl. Phys. {\bf B443} (1995) 85.}
while Type IIA or IIB superstrings compactified on a Calabi-Yau
three-fold get massless charged hypermultiplets when certain
cycles collapse \ref\strominger{A. Strominger,
``Massless Black Holes And Conifolds In String Theory,''
Nucl. Phys. {\bf B451} (1995) 96.}.
The known examples involve special properties of
Ramond-Ramond charges, but it has not been clear that these properties
are essential, and there has been much interest in  whether
similar phenomena can occur  for the heterotic string.

In the known examples, supergravity considerations show that
 the non-perturbative
phenomenon, if it occurs
at all, cannot depend on the value of the string coupling constant.
 For instance, one uses \strominger\
the low energy decoupling of vector multiplets and hypermultiplets
in $N=2$ theories in four dimensions \ref\dewit{B. de Wit,
P. Lauwers, and A. Van Proeyen, ``Lagrangians Of $N=2$
Supergravity-Matter Systems,'' Nucl. Phys. {\bf B255} (1985)
269.}.  Such an explanation
is also possible for the case that we will treat in this paper.
But these phenomena must also be compatible with what one can learn
from conformal field theory, and apparently
 \ref\uwitten{E. Witten, ``Some Comments On String
Dynamics,'' hep-th/9507121.}
from that point of view  the key  is that these
phenomena involve conformal field theory singularities of a special
kind.  The target space loses its compactness, developing a long
tube in which the dilaton grows and the effective string
coupling constant goes to infinity; reducing the string coupling merely
causes the same non-perturbative phenomenon to occur ``farther down the
tube.''

This explanation makes  clear at least one place
where we might look for a similar
phenomenon in the heterotic string.  Associated with
Yang-Mills instantons on
${\bf R}^4$ are $(0,4)$ conformal field theories that mimic
the structure of field theory instantons when the instanton
scale is much greater than the string scale.  But as the instanton
shrinks to zero size, the space-time develops \ref\callan{C. G. Callan,
Jr., J. A. Harvey, and A. Strominger, ``World-brane Actions For String
Solitons,'' Nucl. Phys.  {\bf B367} (1991) 60, ``Supersymmetric String
Solitons,'' in {\it Proc. String Theory And Quantum Gravity '91}
(Trieste, 1991).} an infinite
tube in which the dilaton diverges.  This and related examples
in complex dimension three (discussed in \uwitten\ in mean field
theory) are candidates for situations in which the heterotic
string might exhibit a surprise similar to what has been found
for Type II.

In this paper, we will make a proposal for just what surprise
occurs, at least in the case of the $SO(32)$ heterotic string\foot{For
$E_8\times E_8$, I do not know the answer.}:
the heterotic string develops an extra $SU(2)$ gauge
symmetry when an instanton shrinks to zero size.  After explaining
this result in section two, we check it in section three by comparing
to Type I Dirichlet five-branes
\ref\polch{J. Polchinski, ``Dirichlet-Branes And Ramond-Ramond Charges,''
hep-th/9510017.}
and the ADHM construction of instantons \ref\adhm{M.
F. Atiyah, V. G. Drinfeld, N. J. Hitchin, and Y. I. Manin, ``Construction
Of Instantons'' Phys. Lett. {\bf A65} (1978) 185.}.
In section four, we explain interesting applications of the result
 to the $H$-monopole
problem \ref\gaunt{J. Gauntlett and J. A. Harvey, ``$S$-Duality And
The Spectrum Of Magnetic Monopoles In String Theory,'' hep-th/9407111.} and
the problem of seeing the Type IIA superstring as a heterotic string
soliton \ref\harstrom{J. A. Harvey and A. Strominger, ``The Heterotic
String Is A Soliton,'' Nucl. Phys. {\bf B449} (1995) 535.}.

\newsec{The Heterotic String On ${\bf R}^6\times {\rm K3}$}

To begin with, we consider  the $SO(32)$ heterotic string
compactified on $\R^6\times {\rm K3}$.  The K3 compactification breaks
one half of the supersymmetries; the unbroken supersymmetries
have definite chirality and transform as two spinors of $SO(1,5)$.
Even though this is the smallest possible amount of supersymmetry
in six dimensions, we will call it $N=2$ as it reduces to $N=2$ in four
dimensions.
For a summary of the structure see \ref\sezgin{
A. Salam and E. Sezgin, eds., {\it Supergravity In Diverse Dimensions}
(North-Holland, Amsterdam, 1989).}.

The conformal field theory of the heterotic
string on $\R^6\times {\rm K3}$
is a $(0,4)$ model defined to lowest order in $\alpha'$ by picking
a hyper-Kahler metric on K3, a $B$-field, and a left-moving gauge bundle
with instanton number 24, to cancel anomalies.   It is eventually
these instantons that we will want to study.

The bosons of the theory are arranged in the following supermultiplets.
Vector bosons of unbroken
gauge symmetries are in vector multiplets (which in six dimensions contain
no scalars).  All scalars are in hypermultiplets, with the exception
of the dilaton which is in a ``tensor multiplet'' whose other bosonic
elements are the anti self-dual part of $B$. The bosonic fields in the
supergravity multiplet are the graviton and the self-dual part of $B$.

The $N=2$ supersymmetry is very restrictive and makes certain rather
subtle properties clear.  For instance, while conformal invariance
of $(0,2)$ models defined on Calabi-Yau three-folds is a vexing problem
that is still not well understood, on K3 the situation is easy:
the supersymmetry does not permit the generation of a potential
for the hypermultiplet, which would be the space-time manifestation of
a failure of world-sheet conformal invariance, so these sigma models are
conformally invariant with the expected parameters.\foot{Alternatively,
K3 compactification gives $(0,4)$ supersymmetry, rather than $(0,2)$, and
 leads to conformal invariance because of
considerations involving the $SU(2)_R$ symmetry in the $(0,4)$ algebra
\ref\banks{T. Banks and N. Seiberg, ``Non-perturbative Infinities,''
Nucl. Phys. {\bf B273} (1986) 157.}.}   Likewise, there is
no possible potential involving the dilaton, so these models correspond
to an exact family of quantum vacua.

The hypermultiplets take values in a quaternionic manifold; the structure
and indeed the metric of this manifold are independent of the dilaton,
since $N=2$ supersymmetry does not permit a low energy coupling
of the tensor
multiplet (which in five dimensions would be equivalent to a vector
multiplet) to the hypermultiplets.  This argument, of course, is in
parallel with one in \strominger\ concerning Type II compactification
on a Calabi-Yau manifold.  So singularities in hypermultiplet moduli
space can be found in conformal field theory.

Suppose that in hypermultiplet moduli space we find a
singularity (at finite
distance);
how is it to be interpreted?  Experience has shown \strominger\ that one
should find a physical explanation of such a singularity, and not just
interpret it as a ``singularity in the laws of nature.''
An orbifold singularity might possibly be interpreted simply in terms
of a restored finite gauge symmetry, but
a singularity worse than that should be interpreted in terms
 of some extra supermultiplet going to zero mass.
This supermultiplet should presumably contain only particles of spins
$\leq 1$.  But the only {\it massive} supermultiplet allowed by
six-dimensional $N=2$ supersymmetry is the massive vector multiplet,
which in the $m\to 0$ limit consists of a massless vector multiplet
and hypermultiplet.  For $m\not=0$ there is a Higgs mechanism, with
the vector swallowing one component of the hypermultiplet and gaining
a mass.  In short, the only relevant low energy dynamics allowed by
supersymmetry is that the Higgs mechanism might be turned on or off.
A (non-orbifold) singularity in hypermultiplet moduli space will have
to be interpreted in terms of an enhanced gauge symmetry with extra
massless vector and hypermultiplets.

Though a variety of possible singularities in this
moduli space might be considered, the case that we will look at
is the case in which one of the instantons shrinks to zero size.
One can think of the K3 as being very large so that the instanton
will behave  as an instanton on $\R^4$; the reason to consider
K3 compactification is largely  to embed the instanton physics in a
convenient
global context where some useful deductions (as above) can
be made from supergravity.  Instanton moduli space has a sort of conical
singularity
(even in the differentiable structure, and certainly therefore in
the metric)
when an instanton shrinks to zero size.  This singularity is at a finite
distance.\foot{One can approach a zero size instanton with the instanton
restricted to lie in a given $SU(2)$ subgroup; the metric in this subspace
is determined by the symmetries and quaternionic structure, as in
\gaunt.  There is a simple orbifold
singularity which is at a finite distance.  For larger groups, the
distance
is still finite (as one can restrict to the $SU(2)$ subspace) but the
singularity is worse than an orbifold singularity.}  Therefore,
we should aim to interpret it in terms of extra massless particles present
at the singularity, and so in terms of extra gauge symmetry.

To find what the extra gauge symmetry is -- and how the extra massless
hypermultiplets transform -- we will have to examine the nature of the
singularity associated with a collapsing instanton.  This is done below.
We will get a simple solution in which the extra gauge symmetry
 for a single collapsing instanton is $SU(2)$.  We will not try to prove
that this answer is unique, but it will be clear than an alternative
answer would be far less economical.  In section three, by comparing
to Dirichlet five-branes \polch\
(of Type I superstrings) and the ADHM construction of instantons, we will
argue that when $k$ instantons collapse to the {\it same} point, the
extended gauge group is $Sp(k)$.

As noted in the introduction, from a world-sheet point of view, what
makes possible
these peculiar results -- which arise for arbitrarily weak coupling,
but not in conformal field theory -- is that as an instanton shrinks
to zero, the target space develops an infinite tube in which the dilaton
blows up \callan.  This means that at those particular points in moduli
space where an instanton is shrinking away, perturbation theory loses
its validity for modes supported far down the tube.  Just as in the
conifold
case, the conformal field theory answer may well involve more exotic
behavior (perhaps involving
infinitely many particles propagating down the tube,
as suggested in \uwitten) as opposed to the description with finitely
many light modes that we will propose at the quantum level.

\subsec{Higgs Effect In $N=2$ Supersymmetry In Six Dimensions}

Since the only possible relevant dynamics involves the Higgs effect and
restoration of gauge symmetry, we will first review the Higgs effect in
$N=2$ supersymmetry in six dimensions.  (These remarks are standard
but are included for convenience.)  The Higgs effect
 involves the behavior of
hypermultiplets and gauge multiplets near a point in hypermultiplet moduli
space where part of the gauge symmetry is restored (or broken).
In general, the hypermultiplets take values in a quaternionic
manifold ${\cal M}_0$ which is smooth near the phase transition point;
but after removing supermultiplets that get masses from the Higgs
effect (or can be gauged away after symmetry breaking), one
reduces to a quaternionic manifold ${\cal M}$ -- the moduli space of vacua --
that will have a singularity.  In probing the nature of this singularity,
unrenormalizable interactions like gravity and the curvature of ${\cal M}_0$
are not important; we can turn off gravity and take ${\cal M}_0$ to be
flat.

So we begin with  $k$ hypermultiplets with ${\cal M}_0=\R^{4k}$.
The maximal possible gauge symmetry of the $k$ hypermultiplets
is $Sp(k)$; the action of $Sp(k)$ on $\R^{4k}$ commutes with
the action on the hypermultiplets of the $SU(2)_R$ symmetry
of six-dimensional super Yang-Mills symmetry ($SU(2)_R$ is   explicitly
broken in the string theory by interactions of higher dimension
that will not play a role).  The hypermultiplets transform
as $({\bf 2k},{\bf 2})$ of $Sp(k) \times SU(2)_R$; this is a real
representation as the ${\bf 2k}$ of $Sp(k)$ and the ${\bf 2}$ of
$SU(2)_R$ are both pseudo-real.  If we write the bosonic
part of the hypermultiplets
as $H^{XA}$, where $X,Y=1,\dots ,2k$ and $A,B=1,2$ will be the  $Sp(k)$ and
$SU(2)_R$ indices, then the reality contition on $H$ is
\eqn\really{\bar H_{XA}=\gamma_{XY}\epsilon_{AB} H^{YB}. }
Here $\gamma$ is the invariant antisymmetric tensor of $Sp(k)$
and $\epsilon$ is the invariant antisymmetric tensor of $SU(2)$.
Indices will be raised and lowered using $\gamma$ and $\epsilon$,
so for instance for an $SU(2)_R$ doublet, we write
$v_A=\epsilon_{AB}v^B$, $v^A=\epsilon^{AB}v_B$.

The adjoint representation of $Sp(k)$ consists of symmetric second
rank tensors formed from the fundamental representation.  So
the generators $T^a,\,\,a=1\dots {\rm dim}\,G$
of the gauge group $G\subset Sp(k)$ correspond to symmetric
tensors $T^a_{XY}$ which obey a reality condition $\bar T^{a\,XY}
=\gamma^{XX'}\gamma^{YY'}T^a_{X'Y'}$.
 For every generator, define the $D$-fields
\eqn\eelly{D^a_{AB}= \sum_{X,Y}T^a_{XY}H^X_AH^Y_B.}
Note that these functions obey a reality condition
$\bar D^{a\,AB}=\epsilon^{AC}\epsilon^{BD}D^a_{CD}$.
These functions (which are sometimes called the components
of the hyper-Kahler moment map \ref\rocek{N. J. Hitchin, A. Karlhede,
U. Lindstrom, and M. Rocek, ``Hyperkahler Metrics And Supersymmetry,''
Commun. Math. Phys. {\bf 108} (1987) 535.})
generalize the usual $D$-fields of $N=1$ supersymmetry in four
dimensions. The scalar potential of the theory is
\eqn\juj{V=\sum_{a,A,B}{1\over 2e_a^2}|D^a_{AB}|^2,}
with $e_a$ the gauge couplings.  The classical moduli space ${\cal M}$
of vacua is obtained by setting $V=0$,
that is $D=0$, and dividing by the gauge group $G$.  ${\cal M}$ will
generally be singular when $H$ vanishes  or has any special
value at which the unbroken gauge symmetry is enhanced, because
$G$ does not act freely and because setting to zero the quadratic
functions $D$ generically produces a kind of conical singularity.

Note that if there are $k$ hypermultiplets, and $d$ is the dimension
of $G$, then the dimension of ${\cal M}$ is
\eqn\hub{{\rm dim}\,{\cal M}= 4(k-d).}

\subsec{The One-Instanton Moduli Space And Its Interpretation}

Now we come to the main point of this paper -- the one-instanton
moduli space and the interpretation of the point-instanton
singularity in terms of a Higgs
effect.   As already explained, we consider this problem in the context
of the $SO(32)$ heterotic string on ${\bf R}^6\times {\rm K3}$.
The vacuum is defined among other data by a gauge bundle
with instanton number 24.  We consider vacua for which the instantons
are embedded in an $SO(N)$ subgroup of $SO(32)$ for some fixed
$N$.  We take $N\geq 4$ to avoid special features of small $N$,  but
otherwise the precise value of $N$ is immaterial.
For brevity, we consider the case that the
subgroup of $SO(32)$ left unbroken by the instantons (even when
one of them shrinks away) is precisely
$SO(32-N)$.\foot{$N=24$ is the
largest value that can be achieved by 24 instantons on K3; one can
see this by counting massless hypermultiplets as in \ref\kachru{S. Kachru
and C. Vafa, ``Exact Results For $N=2$ Compactifications Of Heterotic
Strings,'' Nucl. Phys. {\bf B450} (1995) 69.}.
If $N<24$, we are sitting at a somewhat singular point
in the moduli space, but the singularity is decoupled from the fields that
are  $SO(32-N)$ singlets, which are the ones that we will look at.
We look at the the singularity arising in the $SO(32-N)$-invariant
subspace of the moduli space  when an instanton shrinks
to zero.}

When an instanton shrinks to zero size, the details of the K3 are
irrelevant; the instanton may as well be embedded in $\R^4$.  Also,
once an instanton shrinks away, one is left
with an  $SO(N)$ gauge bundle, of instanton number 23, whose
details are irrelevant: it just furnishes a mechanism of completely
breaking the $SO(N)$ symmetry.  The result is that instead of thinking
about instantons on K3, we can simply think about the one-instanton
problem on $\R^4$, with a recipe of {\it not} dividing by global
$SO(N)$ gauge transformations in constructing the moduli space.

Recall then the structure of the one-instanton moduli space on $\R^4$.
The instanton really has structure group $K=SU(2)$, and
is described by its position, scale size, and embedding in $SO(N)$.
For the instanton number to be one,
the embedding must be such that the subgroup of $SO(N)$ commuting
with $K$ is isomorphic to $SU(2)\times SO(N-4)$; thus $SO(N)$
contains $SO(4)\times SO(N-4)\cong SU(2)\times SU(2)\times SO(N-4)$,
 and $K$ must be the first $SU(2)$ factor in such a subgroup.
Specifying the
instanton embedding in $SO(N)$ means telling how  $K=SU(2)$ is embedded
in $SO(N)$, and also giving an ``$SU(2)$ orientation'' of the
instanton, which is a set of three real numbers, acted on transitively
by $SU(2)$, which parametrize how a standard instanton of fixed size
and position can be mapped  into $K$.

To try to   guess a description of the moduli space ${\cal M}$
of these instantons by Higgsing
of a collection of free hypermultiplets, note first that
${\cal M}$  has dimension $4N-8$.  Four of the coordinates are the
instanton center of mass, which certainly decouples from the singularity,
so we write ${\cal M}=\R^4\times {\cal M}'$, where ${\cal M}'$,
the moduli space of instantons centered at the origin, has dimension
$4N-12$.  From \hub, it follows that if there are $k$ hypermultiplets
and the gauge group $G$ has dimension $d$, then $k-d=N-3$.  Moreover,
we will accept at face value a lesson  learned from the behavior
\callan\ of the classical instanton solution: as the instanton shrinks
to zero, it disappears ``down the tube'' and the full $SO(N)$ is restored.
Hence the
$t$ hypermultiplets must form a representation of $SO(N)$.
The most obvious way to obey this condition is to take $N$ hypermultiplets
in the adjoint representation of $SO(N)$ with $G=SU(2)$.
Note that for $N$ hypermultiplets, the gauge group must be
a subgroup of $Sp(N)$.  $Sp(N)$ contains $SO(N)\times SU(2)$, and
the $N$ hypermultiplets are composed of  $2N$ complex fields transforming
as $({\bf N},{\bf 2})$.

It turns out that this choice of gauge group and massless hypermultiplets
leads to the correct moduli space,
while other solutions of the problem
would clearly be far less economical.  Our proposal for the physics
of the $SO(32)$ heterotic string when an instanton shrinks to zero is
thus that the singularity comes from a new $SU(2)$ gauge symmetry
that appears when an instanton shrinks away;
there are moreover 32 massless hypermultiplets, transforming
as $({\bf 32} , {\bf 2})$ under $SO(32)\times SU(2)$.  (Only $N$ of these
figured in the last paragraph as we have been
 discarding the $SO(N)$ singlets.)

To see how the moduli space emerges in setting the $D$-fields
to zero and dividing by $SU(2)$, proceed as follows.  Write the
scalars in the hypermultiplets as $H^{iA'A}$, where now $i=1,\dots, N$
is the $SO(N)$ index, $A'=1,2$ is the $G=SU(2)$ index, and as before
$A=1,2$ carries the $SU(2)_R$ quantum number.  (Thus, we replace the
$Sp(k)$ index $X$ by the pair $iA'$.)  The $D$-field is now
\eqn\ollop{D_{A'B',AB} =\sum_i
\left(H^i_{A'A}H^i_{B'B}+H^i_{B'A}H^i_{A'B}\right)}
and (being in the adjoint representation of both $G$ and $SU(2)_R$)
is symmetric in $A',B'$ and in $A,B$.

We can interpret this formula as follows, using the identification
of $SU(2)\times SU(2)_R$ with $SO(4)$.  The bosonic part of
the hypermultiplets
transform as $({\bf N},{\bf 4})$ of $SO(N)\times SO(4)$, so
in $SO(4)$ language we could write them as $H^{i\alpha}$, where $i=1,\dots,
N$ is the $SO(N)$ index and
$\alpha=1,\dots ,4 $ is the $SO(4)$ index (the components of
$H^{i\alpha}$ are all real).
If we let $V\cong \R^{N}$ be
a copy of Euclidean space transforming in the ${\bf N}$ of $SO(N)$,
then the components  of $H^{i\alpha}$ for fixed $\alpha$ can be
regarded as the components of vectors $h^{\alpha}\in V$, with $\alpha
= 1,\dots, 4$.  The $D$-fields written in
\ollop\ transform as the symmetric traceless tensor of $SO(4)$.
The condition that $D=0$ becomes in this language
\eqn\ollp{(h^\alpha,h^\beta)={\rm constant}\cdot \delta^{\alpha\beta},}
where $(~,~)$ is the inner product in $V$.

This means that after division by a constant $\rho$, the $h^\alpha$,
$\alpha=1,\dots, 4$ are orthonormal vectors.  $\rho$ is called
the instanton scale size.  The choice of the four orthonormal vectors
$h^\alpha\in V$    breaks $SO(N)$ down to $SO(N-4)$, but
after dividing by $G=SU(2)$, the unbroken subgroup of $SO(N)$
is $L= SU(2)\times SO(N-4)$, where the $SU(2)$ is unbroken up to a $G$
transformation.  The subgroup $K$ of $SO(N)$ that commutes with
$L$ is isomorphic to $K=SU(2)$ and is the gauge
group of the instanton.  $K$ only depends on the four-dimensional
subspace $B\subset V$ spanned by the orthonormal vectors $e^\alpha=
h^\alpha/\rho$.  Once $B$ is fixed, the choice of basis vectors $e^\alpha$ of
$B$ depends (after dividing by $G$) on three real parameters, which correspond
to how the instanton is oriented in $K$.  All this agrees precisely
with the structure of the one-instanton moduli space.

So we get a consistent picture of the behavior in the limit of a small
instanton.

\subsec{Comparison To Conformal Field Theory And Generalization}

Let us now make a few remarks aimed at underscoring that the phenomenon
we have found cannot be seen in conformal field theory.
We start with the $SO(32)$ heterotic string with a gauge bundle of
instanton number 24.  The most extreme case of the phenomenon considered
above is that all 24 instantons might shrink away, restoring the full
$SO(32)$ gauge symmetry and giving us an extra unbroken gauge symmetry
$SU(2)^{24}$ -- one $SU(2)$ for   each collapsed instanton --
assuming that the instantons collapse at distinct points
so that the above analysis suffices.  (From what we will find in the
next section,
the gauge group is $SO(32)\times Sp(24)$ if all instantons collapse at the
same point.)   Now, the heterotic string in six dimensions cannot
possibly have a gauge symmetry $SO(32)\times SU(2)^{24}$ in conformal
field theory, because this would imply a central charge of at least
40, which is much too much.  So the phenomenon is quantum mechanical,
out of reach of conformal field theory.

We formulated the analysis in terms of compactification on
$\R^6\times {\rm K3}$ because -- apart from being an interesting context
for applying the result -- this made it possible to deduce
immediately from supergravity a few useful preliminaries stated
at the beginning of this section.  But the local nature of the question
means that it must carry over when a Yang-Mills instanton (as a function
of four of the variables, independent of the other six) is embedded
on any ten-manifold.  Such an instanton has been interpreted as
a heterotic string solitonic five-brane \ref\strom{A. Strominger, ``Heterotic
Solitons,'' Nucl. Phys. {\bf B343} (1990) 167.},
the idea being that the five-brane world-volume is the codimension-four
manifold on which   the instanton is localized.
In this interpretation, the instanton
scale size is a kind of variable thickness of the five-brane.
Our result means that in the limit of zero thickness, the five-brane
carries a previously unknown $SU(2)$ gauge symmetry along with the massless
six-dimensional hypermultiplets described above.

\newsec{Comparison With Dirichlet Five-Branes And The ADHM Construction}

\nref\dab{A. Dabholkar, ``Ten Dimensional Heterotic String As A Soliton,''
Phys. Lett. {\bf B357} (1995) 307.}
\nref\hull{C. M. Hull, ``String-String Duality In Ten Dimensions,''
Phys. Lett. {\bf B357} (1995) 545.}
\nref\pw{J. Polchinski and E. Witten, ``Evidence For Heterotic - Type I
String Duality,'' hep-th/9510169.}
\nref\tseytlin{A.  Tseytlin, ``On SO(32) Heterotic - Type I Superstring
Duality In Ten Dimensions,'' hep-th/9510173.}

The discussion in the last section involved, in spirit, a weak coupling
analysis of the heterotic string.  But the phenomenon studied is
independent of the coupling, so one can try to test the results by
going to strong coupling and using heterotic - Type I duality
\refs{\witten,\dab - \tseytlin} to convert the problem to a weak
coupling problem involving Type I.  To follow that line of thought,
we need to understand what small instantons do in the Type I theory.

What makes this feasible is a remarkable recent observation
\polch\ that gives a new way to study the small scale size limit
of an instanton.  Note that for Type I, the action of a Yang-Mills
instanton scales with the string coupling $\lambda$ as $1/\lambda$
(because the gauge action comes from the disc, whose contribution is of
order $1/\lambda$).  The instanton thus corresponds to a solitonic five-brane
(in a sense recalled at the end of the last section) with tension
of order $1/\lambda$.  But the  Dirichlet five-brane
\polch\ is another Type I five-brane with a tension of the same order.
The Dirichlet five-brane appears classically to have zero thickness, so
one can ask whether it actually coincides with the zero thickness limit
of the solitonic five-brane, that is, the instanton.

The effort to implement this idea will proceed as follows.
Upon quantizing the Dirichlet five-brane, one does not find world-volume
gauge fields unless one puts in Chan-Paton factors without an immediate
world-sheet explanation.  When this is done,
the desired massless hypermultiplets automatically appear.
Where one really gets a payoff is when one considers $k$ instantons
that simultaneously collapse at the same point, corresponding to $k$
nearly coincident five-branes.  Once Chan-Paton factors are chosen
for a single five-brane, the factors carried by $k$ coincident
five-branes are uniquely determined, and one gets a prediction
for the structure of the Yang-Mills $k$-instanton moduli space on $\R^4$.
This prediction is correct; it agrees with what is known from the ADHM
construction of instantons.  This success convinces me -- and hopefully
the reader -- that the Dirichlet five-brane has the necessary Chan-Paton
factors, which hopefully will have a more complete
world-sheet explanation (beyond a hint that appears below) in due course.

\subsec{Quantization Of The Five-Brane}

We consider in $\R^{10}$, with coordinates $x^0,\dots,x^9$, a five-brane
located at $x^6=\dots = x^9=0$.  Open strings in the field of the five-brane
are permitted to either have ordinary free (Neumann) boundaries, or
to have Dirichlet boundaries located on the five-brane world-volume, that
is at $x^6=\dots = x^9=0$.  The excitations of the five-brane are open
strings that either have two Dirichlet ends (DD) or one end of each
kind (DN).  Quantization of the DD and DN sectors is similar to the
discussion of the Dirichlet string in \pw, which the reader may find useful
for background.

First we consider the DD sector, using covariant RNS formalism.
As in the case treated in \pw,
the ground state energies and mode expansions are almost the
same as for conventional open strings, so the massless states form
a  conventional Yang-Mills vector multiplet, with two differences.
(1) The zero modes of $X^6,\dots,X^9$ are eliminated by the boundary
conditions, so the massless modes are functions of $x^0,\dots,x^5$ only
and can be decomposed in six-dimensional terms as a vector $A_i(x^j)$,
$i,j=1\dots 6$, and scalars $\phi_s(x^j)$, $s=6,\dots , 9$, $j=0,\dots,5$.
(2) Because of the unusual boundary conditions at the ends, the
operator $\Omega$ that exchanges the two ends of the string acts
in an unusual fashion.  The little we need to know follows from the fact
that the vector has a conventional vertex operator $V_A=A_i\,\partial X^i/
\partial \tau$ with $\partial_\tau$ the derivative along the
boundary, while (as in \ref\oldpolch{J. Polchinski,
``Combinatorics of Boundaries In String Theory,'' Phys. Rev. {\bf D50}
(1994) 6041.})
the scalar vertex operator is $V_\phi=\phi_s \partial X^s/\partial \sigma$
with $\partial_\sigma$ the normal derivative.
$V_A$ is odd under reversing the orientation of the boundary while
$V_\phi$ is even.

The last statement means that in the absence of Chan-Paton factors,
when one projects onto states invariant under $\Omega$,
the vector is removed from the spectrum, while the scalar survives.
The scalar indeed represents the fluctuations in the position of the
five-brane.  If we want the five-brane to carry $SU(2)$ gauge fields,
we need to introduce $SU(2)$ Chan-Paton factors.

\bigskip\noindent
{\it Review Of Chan-Paton Factors}

Let us therefore make a brief review of Chan-Paton factors.
These originate in the possibility that the end of the string carries
a charge in some representation $R$ of a gauge group $G$.
Conditions of factorization show \ref\marcus{N. Marcus and A. Sagnotti,
``Tree Level Constraints On Gauge Groups For Type-I Superstrings,''
Phys. Lett. {\bf 119B} (1982) 97.}
that $G$ must be $U(n)$, $SO(n)$, or $Sp(n)$, with $R$ being the fundamental
representation in each case.

For {\it unoriented} strings, the only possibilities are $SO(n)$ and $Sp(n)$.
In terms of world-sheet path integrals, the meaning of the Chan-Paton
factors, for strings propagating in a space-time with gauge field
$A_i(x)$, is that the path integrand contains a factor
\eqn\jugg{\Tr_R  P\exp\oint_C A_i(X(\tau))dX^i(\tau)}
for each boundary  component $C$.  For the strings to be unoriented
means that this factor must be invariant under reversal of orientation of
$C$, which exchanges $R$ with its dual.  Hence $R$ must be real or
pseudoreal, so $SO(n)$ and $Sp(n)$ are possible from this point of view
and $U(n)$ is not.

Now let us determine the $SO(n)$ or $Sp(n)$ quantum numbers of the
scalars $\phi$, in case Chan-Paton factors are introduced.
Since in the DD
sector there is a charge at each end of the string, the $\phi$'s
(or any DD states)  transform as a  second rank tensor in the fundamental
representation.  The question is whether that tensor is symmetric
or antisymmetric.
Since $V_\phi$ contains the normal derivative
instead of the tangential derivative to the boundary -- which gives it
an extra minus sign under orientation reversal relative to $V_A$ --
the $\phi$'s are in each case tensors of the opposite kind relative to the
$A$'s.  For $SO(n)$, the adjoint representation is the  antisymmetric
tensor, so the $\phi$'s transforms as symmetric tensors $\phi_{XY}=\phi_{YX}$,
with $X,Y$ the gauge indices.  For
$Sp(n)$, the adjoint representation
is the {\it symmetric} tensor, so the $\phi$'s transform as {\it antisymmetric
tensors} $ \phi_{XY}=-\phi_{YX}$.

For the basic five-brane of instanton number one, we want the gauge group
to be $SU(2)$.  Luckily this coincides with $Sp(1)$, and so is one of the
allowed cases.  The antisymmetric second rank tensor of $Sp(1)$ happens
to be the trivial one-dimensional representation, so with these
particular Chan-Paton factors, the $\phi$'s are ordinary scalars that
represent fluctuations in the center of mass position of the five-brane
(or instanton) and nothing else.

In the rest of this section, we will see that the Dirichlet five-brane
seems to work perfectly if one endows it with $SU(2)$ Chan-Paton factors.

\bigskip\noindent
{\it Hypermultiplets}

The treatment of the small instanton in section two required, in addition
to $SU(2)$ gauge fields, hypermultiplets propagating on the five-brane
that transform as $({\bf 32},{\bf 2})$ of $SO(32)\times SU(2)$.  They
must come from the DN sector as we did not find them in the DD sector.
This works without any trouble.  The N end of the DN string carries
conventional Chan-Paton factors transforming as ${\bf 32}$ of $SO(32)$, while
the D end carries the new Chan-Paton factors transforming as ${\bf 2}$ of
$SU(2)$, so the entire DN sector transforms as $({\bf 32},{\bf 2})$.

To see that the ground state has the space-time structure of
a massless hypermultiplet, look at the bosons from the Neveu-Schwarz
sector.  In light cone gauge, the bosons $X^2,\dots, X^5$ have the
usual  mode expansion with integer modes, while $X^6,\dots,X^9$ have
an expansion in half-integer modes because of an extra minus sign in
reflection at the D boundary.  For fermions, the situation is opposite:
$\psi^2,\dots,\psi^5$ have the usual half-integer modes of the Neveu-Schwarz
sector, while $\psi^6,\dots,\psi^9$ have an expansion in integer modes
because of the extra minus sign in reflection at the D boundary.
Since the same number of bosons and fermions have boundary conditions
of each kind, the ground state energy is zero.  The ground state
degeneracy comes from quantizing the zero modes of $\psi^6,\dots,\psi^9$.
Their quantization gives four states, all of spin zero under rotations
of $x^0,\dots,x^5$.  Two of the four survive the GSO projection.
The supermultiplet whose bosonic part consists of two scalars is the
hypermultiplet, so we have shown that massless  DN excitations are
a single $({\bf 32},{\bf 2})$ of hypermultiplets.

Note that the two bosonic states just discussed transform as a definite
chirality spinor
under the $SO(4)$ rotation group of $x^6,\dots,x^9$.
This representation is pseudoreal, and while {\it two-dimensional}
in the complex sense, it is {\it four-dimensional} when viewed over the
real numbers.  Since bosons are real (that is, the real and imaginary
parts of a complex bose field are independent degrees of freedom), one
should really count the real dimension, and therefore in the absence
of Chan-Paton factors, the pseudoreal nature of this particular representation
would force a peculiar doubling of the spectrum, which does not seem to have
been encountered before in string theory.  The $SU(2)$ Chan-Paton factors
give a natural interpretation of this doubling, a fact which may even
contain the germ of an explanation of why the Chan-Paton factors are present.

\subsec{Predicting The Moduli Space Of $k$-Instantons On $\R^4$}

A strong confirmation of this picture comes when we consider $k$ instantons
that simultaneously coalesce and collapse.  The key point is to understand
the case of $k$ point instantons at the same point in space-time.  This
is the most degenerate case that will have the largest extended gauge
symmetry.  Once the gauge group and massless hypermultiplets are understood
in this case, the general configuration is obtained by the Higgs mechanism
-- letting the hypermultiplets get expectation values, subject to vanishing
of the $D$-fields.

We take our instanton solutions  to be functions of $x^6,\dots,x^9$,
and we consider the case in which the instanton collapses to a point.
A single point instanton corresponds to the Dirichlet five-brane discussed
above.  $k$ point instantons correspond therefore to $k$ parallel
D-branes.  Each D-brane carries its own $SU(2)$ gauge symmetry, due
to open strings starting and ending on that D-brane.  With $k$ D-branes,
we must also consider open strings starting on one D-brane
and ending on another; these transform as doublets under the two  $SU(2)$'s
of those two D-branes  and singlets of the
others.  The ground state in such a sector (as in a similar case considered
in \ref\bound{E. Witten, ``Bound States Of Strings And $p$-Branes,''
hep-th/9510135.},
where the point is explained more fully)
is a vector multiplet with a mass proportional to the distance from the
starting point to the end-point.  In the limit as all of the
D-branes coincide, these states all become degenerate, making up the
vector multiplets of an unbroken $Sp(k)$ gauge symmetry.  This then
must be the gauge group for $k$ coincident point instantons.  We have
simply learned that -- as one might guess intuitively -- if a single
five-brane has $Sp(1)$ Chan-Paton factors, $k$ coincident ones have
$Sp(k)$ Chan-Paton factors.

The massless hypermultiplets in this configuration are simply what one
gets if one quantizes the five-brane with the $Sp(k)$
Chan-Paton factors.  So from our above
discussion, the massless hypermultiplets from the DD
sector transform as the antisymmetric product of two copies of the
fundamental representation of $Sp(k)$ (and are $SO(32)$ singlets,
of course), while those from the DN
sector transform as $({\bf 32},{\bf 2k})$ under $SO(32)\times Sp(k)$.
Note that this representation is reducible, decomposing as the direct
sum of a singlet (the multiples of the invariant antisymmetric tensor
of $Sp(k)$) and the ``traceless antisymmetric tensors.''  In the remarks
that follow, the expectation value of the   singlet hypermultiplet
corresponds to the center of mass of the $k$-instanton system.

To get a general $k$-instanton configuration, one simply
lets the hypermultiplets get vacuum expectation values (with the usual
constraints and gauge invariance).  Thus we get a prediction for the
moduli space of $k$-instanton solutions of $SO(32)$ gauge theory on
$\R^4$: it is the moduli space of vacua of the six-dimensional
$Sp(k)$ gauge theory with the massless hypermultiplets found
in the last paragraph.  This prediction is correct; that is, it agrees
with what is proved using the ADHM construction of instantons.

The ADHM construction \adhm, in other words, determines the actual
instanton solutions, and gives as a byproduct a description of the instanton
moduli space.
The instanton moduli space is obtained as a hyper-Kahler quotient
or -- in more physical terms -- as the moduli space of vacua of
an $N=2$ theory with a certain collection of vector multiplets
and hypermultiplets.
While originally formulated for gauge group $Sp(N)$,
the ADHM     construction can be adapted to $SU(N)$ or $SO(N)$
by imposing invariance conditions that reduce $Sp(N)$ to $SU(N)$ or
$SO(N)$.  The relatively  recent treatments,
such as \ref\goddard{E. Corrigan and P. Goddard, ``Construction Of
Instanton And Monopole Solutions and Reciprocity,'' Ann. Phys. {\bf 154}
(1984) 253.}, have most frequently focused on $SU(N)$.   One source
where $SO(N)$ is treated and
one can find the description we have obtained
of the  $SO(N)$ multi-instanton moduli space is
\ref\christ{N. H. Christ, E. J. Weinberg,
and N. K. Stanton, ``General Self-Dual Yang-Mills Solutions,''
Phys. Rev. {\bf D18} (1978) 2013.}.  The hypermultiplets that
we obtained from the DD and DN sectors correspond, respectively, to
$b^r$ and $v$ in equation (4.11) of \christ, while what in that
reference is called the ``reality condition'' on $b^r$ and $v$ is
the vanishing of the $D$-fields.

\newsec{Soliton Strings And $H$-Monopoles}

In this concluding section, we will see how these results are relevant
to two of the striking puzzles in the recent literature, as follows.

\nref\duffs{M. J. Duff and J. X. Lu, ``Loop Expansions And
String/Fivebrane Duality,'' Nucl. Phys. {\bf B357} (1991) 534;
M. J. Duff and R. R. Khuri, ``Four-Dimensional String/String Duality,''
Nucl. Phys. {\bf B411} (1994) 473; M. J. Duff and J. X. Lu,
``Black And Super $p$-Branes In Diverse Dimensions,'' Nucl. Phys.
{\bf B416} (1994) 301; M. J. Duff and R. Minasian, ``Putting
String/String Duality To The Test ,'' Nucl. Phys. {\bf B436} (1995) 507;
M. J. Duff, R. R. Khuri, and L. X. Lu, ``String Solitons,''
Phys. Rep. {\bf 259} (1995) 213; M. J. Duff, ``Strong/weak
Coupling Duality From The Dual String,'' Nucl. Phys. {\bf B442} (1995)
47.}
(1) It is    believed that the heterotic string on $\R^6\times \T^4$ is
equivalent to the Type IIA theory on $\R^6\times {\rm K3}$.
The low energy field theories match up in such a way that an
elementary string in $\R^6$ in one of the two theories must correspond
to a soliton in the other \refs{\duffs,\witten}.  On each side there is
a soliton string that is a candidate for corresponding to the elementary
string on the other side.  The soliton strings have the correct
current-carrying ability in each case \ref\sen{A. Sen,
``String String Duality Conjecture In Six Dimensions And Charged
Solitonic Strings,'' Nucl. Phys. {\bf B450} (1995) 103.}
but if one considers more precisely the quantization \ref\harstrom{J.
A. Harvey and A. Strominger,  ``The Heterotic String
Is A Soliton,'' Nucl. Phys. {\bf B449} (1995) 535.},
one finds
that the soliton string of Type IIA beautifully reproduces the elementary
heterotic string, but the soliton string of the heterotic string
seems to lack the zero modes needed to reproduce the structure
of the elementary Type IIA string.

(2) There is by now   overwhelming evidence for $S$-duality in the
heterotic string toroidally compactified to four dimensions, yet the
first attempt at a really stringy  test of this symmetry,
which was the search for the $S$-duals of certain magnetic monopoles
\ref\gauntharv{J. Gauntlett and J. A. Harvey,
``$S$-Duality And The Spectrum Of Magnetic Monopoles In Heterotic
String Theory,'' hep-th/9407111.}, led to an apparent
contradiction, again because the necessary zero modes were absent.

The two problems are quite closely related because both involve the
behavior of small Yang-Mills instantons in the heterotic string.
\foot{Previous authors have worked in special vacua with unbroken non-abelian
gauge symmetry so that big instantons contribute as well as the small
ones that we will meet below, but this seems only to add infrared
problems to the unavoidable subtleties from small instantons.}
I will not give a complete solution of either problem, but we will
get far enough, I believe, to make it clear that the gauge symmetry
of the five-brane is an essential part of the story.

\subsec{Soliton Strings}

\bigskip\noindent
{\it Preliminaries}

First of all, to what extent, by studying a soliton string, can one
expect to learn anything about string-string duality?

Mere existence of a soliton string is not convincing evidence that the
string becomes elementary in some regime, any more than mere discovery
of a soliton in field theory is evidence that there is a regime in
which it is elementary.  In field theory, a soliton can be usefully
treated as elementary only in a regime in which (i) it becomes
much lighter than the natural mass scale of the theory, and (ii) it is
weakly coupled.  Condition (i), in particular, is very special.
For BPS-saturated solitons in a supersymmetric theory, the fact that the
mass goes to zero can sometimes be predicted from the structure of the
central charges \ref\wo{E. Witten and D. Olive, ``Supersymmetry Algebras
That Include Central Charges,''  Phys. Lett. {\bf 78B} (1978) 97.},
but this can be quite delicate.
For $N=2$ theories in four dimensions, for instance, such an argument
requires detailed knowledge of the dynamics \ref\sw{N. Seiberg and
E. Witten, ``Electric-Magnetic Duality, Monopole Condensation,
And Confinement In $N=2$ Supersymmetric Yang-Mills Theory,''
Nucl. Phys. {\bf B426} (1994) 19.},
while for $N\geq 4$ in four dimensions, or with
enough supersymmetry in other dimensions, the relevant low energy
dynamics is trivial, and an argument of this type is much more robust.

For a soliton string to become elementary, (i) its tension must go to zero
in Planck units, and (ii) it must become weakly coupled and contain
the graviton as one of its states.  (There is some redundancy in this
formulation since if the tension is small, the gravitational interactions
are weak, and if moreover the graviton is a string state, that indicates
that the string coupling is small.)  With present techniques, the main
way to verify (i) is to use a BPS formula.  In the case of comparing
the heterotic string on $\R^6\times \T^4$ to Type IIA on $\R^6\times {\rm
K3}$, the six-dimensional $N=4$ supersymmetry prevents modifications of
the BPS formula seen at tree level (roughly as in
\ref\dabha{A. Dabholkar and J. A. Harvey, ``Non-Renormalization
Of The Superstring Tension,'' Phys. Rev. Lett. {\bf 63} (1989) 478}),
so it is reasonable to claim
that the respective soliton strings, which have tensions of order
$1/\lambda^2$ (with $\lambda$ the string coupling) go to zero tension
for strong coupling.  As for condition (ii), it is harder to verify,
but the comparison of the soliton string of Type IIA to the
elementary heterotic string \harstrom\ is so striking as to strongly
suggest that condition (ii) is also true in that case.  (For an example
in which (i) is true but not (ii), see \uwitten.)

Before attempting a similar analysis in the reverse direction, let
us discuss just how precise a result one should hope for.  To discuss
soliton strings, it is usually convenient to replace $\R^6$ by
$\R^5\times \S^1$,
with a large circumference for $\S^1$, and consider the soliton string
to wrap once around $\S^1$.  The best result one could hope for is then
to show a natural one-to-one correspondence between states of the soliton
string on one side and states of the elementary string on the other side.
This was achieved by Harvey and Strominger in comparing the quantization
of the string soliton of Type IIA to the elementary heterotic string.

This, however, is a much stronger result than one necessarily needs or
should expect in other cases.  Even if a soliton string of string theory
A does for {\it strong coupling} turn into an elementary string of string
theory B, it does not follow that the quantization of the theory A soliton
for {\it weak coupling} will match up with elementary string B.
In general, one might expect such a matching only for the BPS-saturated states,
that is, for those string winding states that are BPS-saturated and
so form small supermultiplets that can (when there is enough supersymmetry)
be followed reliably from weak to strong coupling.
\nref\ellgen{A. N. Schellekens and N. P. Warner, ``Anomalies And Modular
Invariance In String Theory,''
Phys. Lett. {\bf 177B} (1986) 317, ``Anomaly Cancellation And
Self-Dual Lattices,'' Phys. Lett. {\bf 181B} (1986) 339,
``Anomalies, Characters, and Strings,''
Nucl. Phys. {\bf B287} (1987) 317.}
\nref\wittene{E. Witten, ``Elliptic Genera And Quantum Field Theory,''
Commun. Math. Phys. {\bf 109} (1987) 525, ``The Index Of The Dirac
Operator In Loop Space,'' in {\it Elliptic Curves And Modular Forms
In Algebraic Topology}, ed. P. Landwebber (Springer-Verlag, 1988).}
\nref\lerche{W. Lerche, ``Elliptic Index And Superstring Effective
Action,'' Nucl. Phys. {\bf B308} (1988) 102.}
\nref\eguchi{T. Eguchi, H. Ooguri, A. Taormina, and S. K. Yang,
``Superconformal Algebras And String Compactification On Manifolds
With $SU(n)$ Holonomy,'' Nucl. Phys. {\bf B315} (1989) 193.}
If then we want to see the elementary Type IIA string on $\R^6\times {\rm K3}$
as a soliton of the heterotic string on $\R^6\times \T^4$, we might first
ask what are the BPS-saturated states.  To be more precise, what are the
BPS-saturated states of
a Type IIA string that wraps once around the $\S^1$ in $\R^5\times \S^1\times
{\rm K3}$?  Here a subtlety arises: we must decide how much BPS saturation
we want.
With $\bar L_0$ and $L_0$ the left and right-moving world-sheet Hamiltonians,
BPS-saturation for right-movers means $L_0=0$, and that for left-movers
means $\bar L_0=0$.  BPS-saturation for both left and right-movers gives
therefore $L_0=\bar L_0=0$; only finitely many states (associated with
the cohomology of K3) obey these conditions.  If, however, we impose
only $L_0=0$, but not $\bar L_0=0$, the answer is completely different:
the left-moving (or right-moving)
$N=4$ superconformal algebra creates an exponential tower of states
of $L_0=0$ (or $\bar L_0=0$).  Actually,
on K3, there is a full stringy spectrum of states of $L_0=0$
 with a degeneracy
corresponding to four left-moving bosons and four left-moving
fermions (and similarly for $\bar L_0=0$ and right-movers).
This is obvious for K3 orbifolds, and follows more generally
by considering
 the ``elliptic genus'' \refs{\ellgen - \eguchi},
 which is simply the partition function with left-moving  Neveu-Schwarz
and right-moving Ramond boundary conditions, and is computable (as in
\eguchi\ in the K3 case) because of its topological invariance.
These states
fit into ``middle-sized'' supermultiplets, and just like the fully BPS
saturated states in the ``small'' multiplets, can be followed continuously
from weak to strong coupling.  Thus a soliton string that would
reproduce the elementary Type IIA string on K3 must have four left and
right-moving world-sheet bosons and fermions to reproduce the spectrum
of half-saturated states of the K3 conformal field theory; additional
massless world-sheet
fields are needed, of course, to describe the motion in $\R^6$.

\bigskip\noindent
{\it The Instanton As A Soliton String}

We now consider the actual structure of the solitonic string
solution of the heterotic string on $\R^6\times \T^4$ and the attempt
to compare it to the elementary Type IIA string on $\R^6\times {\rm K3}$.

First of all, for the heterotic string on $\R^6\times \T^4$ we pick
a generic vacuum, in which $SO(32)$  is broken to $U(1)^{16}$ by
Wilson lines (and the overall gauge group is $U(1)^{24}$).
On $\R^6$ we pick coordinates $x^0,x^1,\dots,x^5$, with
$x^0$ being the time and the others being space coordinates.
We replace $\R^6$ by $\R^5\times \S^1$, taking $x^5$ to be a periodic
variable.  Now, the analysis of the low energy field theories
shows \refs{\duffs,\witten}  that for a soliton string that has the
quantum numbers of the elementary Type IIA string (on $\R^6\times{\rm K3}$),
the heterotic string description will involve
an instanton in the variables $x^1,\dots,x^4$.
This gives the right quantum numbers, but  we must find a solution
of the heterotic string
with those quantum numbers.  Because of the symmetry  breaking to
the abelian group $U(1)^{24}$
and the fact that there are no abelian instantons on $\R^4$, the
instanton will  classically  shrink to a point (giving the so-called
neutral five-brane of the heterotic string \callan).
So we meet precisely the problem studied in this paper: the behavior
of a point instanton.

Suppose that the point instanton is localized at $x^1=\dots = x^4 = 0$.
This corresponds to a five-brane whose world-volume is
$\R\times \S^1\times \T^4$, with $\R$ being parametrized by $x^0 $ and  $\S^1$
by $x^5$; the $\T^4$ is the original $\T^4$
of the heterotic string compactification.
 The massless world-sheet spectrum of this zero-width five-brane
 was the subject of section two, with the result being that
there is a massless neutral  hypermultiplet,  an $SU(2)$ vector multiplet,
and hypermultiplets with $SO(32)$ quantum numbers.
The $SO(32)$ non-singlets, because of interaction with the Wilson
lines on $\T^4$, lack zero modes
and do not contribute to states with $L_0=0 $ or $\bar L_0=0$.
 The massless neutral hypermultiplet
(which entered as the instanton center of mass or in quantizing
the DD sector of the Dirichlet five-brane) describes the oscillations in
$x^1,\dots,x^4$.  These modes
describe motion of the elementary Type IIA string in $\R^6$.
  The $SU(2)$ vector multiplet will have to give the
modes associated with the K3 conformal field theory.

Since the $\S^1$ is to be considered very big (its finite size
is just a device to avoid messy boundary conditions in discussing
soliton strings), in working on $\R\times \S^1\times \T^4$ we are
essentially compactifying the five-brane from six to two dimensions
on $\T^4$.  The zero modes of the $SU(2)$ vector multiplets on $\T^4$ are
simply the $SU(2)$ flat connections -- whose fluctuations then
propagate as massless world-sheet fields of the soliton string
on $\R\times \S^1$.  A flat connection has four commuting Wilson lines
$W_i$, $i=1,\dots, 4$, representing the holonomies in $x^1,\dots,x^4$.
As they commute, they can be simultaneously diagonalized,
$W_i={\rm diag}(e^{ia_i},e^{-ia_i})$, with the $a_i$ being angles.
The Weyl group acts by exchanging the two eigenvalues, $a_i\to -a_i$.
Thus the moduli space of flat connections is an orbifold
$Y=(\S^1)^4/\Z_2$.

Since $Y$ is a kind of singular K3 orbifold, it seems that we have found
the desired K3 sigma model on the soliton string world-sheet so that
 quantization of the heterotic string soliton  agrees with
the elementary Type IIA string.   But a few points require further
clarification (which will not be given here).
(1) We seem to have obtained a K3 target space, but the ``wrong'' one
(as the K3 on the Type IIA side is not necessarily an orbifold).
Is this somehow corrected to the ``right'' K3 if the quantization is carried
out more precisely, or is it all right
to have the ``wrong'' K3, because of topological invariance of the
BPS half-saturated spectrum?  (2) Do the $\Z_2$
fixed points in $Y$ -- where $SU(2)$ gauge symmetry is restored --
really behave as orbifold fixed points in elementary
string theory?  This is implicit in identifying $Y$ as a K3 of any kind,
even the ``wrong'' one.  (3)  We ignored here the $SO(32)$ non-singlet
hypermultiplets, because the Wilson lines that broke $SO(32)$ to
an abelian group appear to keep them from contributing to the states
with vanishing $L_0$ or $\bar L_0$.  But at certain points in $Y$, for
some of these states, the $SU(2)$ Wilson lines cancel the effects
of the $SO(32)$ Wilson lines;  how does this affect the spectrum?

\subsec{$H$-Monopoles}

The problem of $H$-monopoles arises in quantization of the heterotic
string on $\R^4\times \T^6$ in a generic vacuum with the gauge group
broken to an abelian group by Wilson lines.  We take the $\R^4$ coordinates to
be $x^0$ (time) and $x^1,\dots, x^3$ (space), while the $\T^6$ coordinates
are $x^4,\dots,x^9$.  Consider a string winding once around the $x^4$
direction.  If the right-moving oscillators are placed in their ground
state, the spectrum consists of 24 BPS-saturated $N=4$ vector multiplets.
The number 24 arises because to set $L_0=\bar L_0$, one needs a
single excitation of any one of the 24  left-moving oscillators (in the
bosonic formulation of the heterotic string).  In light cone gauge,
the 24 oscillators (and likewise the 24 vector multiplets)
can be divided into (a) eight which transform as a vector of the $SO(8)$
rotation group of the transverse dimensions (of course, this group is
broken to a subgroup by the compactification); (b) sixteen which are $SO(8)$
singlets.

These winding states are electrically charged with respect to  a
four-dimensional $U(1)$ gauge field which is simply the components
$B_{\mu 4}$, $\mu=0,\dots,3$ of the usual $B$-field.  The model is
believed to have $S$-duality \ref\senschwarz{A. Sen and J. H. Schwarz,
``Duality Symmetries Of       4-D Heterotic Strings,''
Phys. Lett. {\bf B312} (1993) 105.},
in which case there must also be 24
BPS-saturated vector multiplets, magnetically charged with respect
to the same $U(1)$, and  with the same decomposition as above.

The magnetically charged states in question are simply the instantons
in the variables $x^1,\dots,x^4$ -- or at least (by virtue of the
anomaly-canceling equation $dH=\tr F\wedge F$, which can be used
to show that these instantons carry magnetic charge) these are natural
examples of magnetically charged states.
The problem of $H$-monopoles is that quantization of the instantons
 with their standard zero modes does not  give the desired spectrum with
24 degenerate multiplets.

As in our discussion of soliton strings, the problem depends on
understanding the behavior of small instantons, since in a generic
vacuum with the gauge group broken to an abelian subgroup, the instantons
 shrink classically to points.  Let us re-examine the question in light
of the gauge symmetry of the zero size instanton, which gives a richer
than previously expected world-volume structure to the relevant  fivebrane.

The massless world-volume multiplets of the five-brane  are a neutral
hypermultiplet $\phi$, an $SU(2)$ vector multiplet, and hypermultiplets
transforming as $({\bf 32},{\bf 2})$ of $SO(32)\times SU(2)$.  $\phi$,
which represents oscillations in $x^1,\dots , x^4$, has certainly
been part of previous discussions of this problem, so to get a degeneracy
that has been previously lacking we must focus on the others.
The BPS-saturated states we want correspond to supersymmetric ground
states of the five-brane quantization, so the $({\bf 32},{\bf 2})$
hypermultiplets, which generically have no zero modes because of the
Wilson lines, can be neglected at least in a first approximation.
It remains to look at the vector multiplet.

The vacua of the $SU(2)$ vector multiplet are spanned (as in
the discussion of soliton strings) by commuting Wilson lines
that we can take to be $W_i={\rm diag}(e^{i\alpha_i},e^{-i\alpha_i})$
with $i=5,\dots,9$.  This description of the low energy dynamics
is only valid for generic values of the $\alpha_i$ and can fail in either
of two ways.  (1) There are 32 singularities where all $W_i$ are
$\pm 1$, so that $SU(2)$ is restored.  (2) There are
sixteen additional singularities
when the $W_i$ cancel the effects of the $SO(32)$ Wilson lines for
some charged modes so that some of the non-singlets that we have
ignored do have zero modes.

For generic $W_i$, the $SU(2)$ gauge symmetry
of the five-brane world-volume is spontaneously broken to $U(1)$ and
only the $U(1)$ vector multiplet remains massless.  Even at low energies
and ignoring the singularities, the $U(1)$ vector multiplet cannot
be treated simply as free; one must remember the Weyl transformation
$w:\alpha_i\to -\alpha_i$.

If we ignore the singularities and treat the problem as motion on
a moduli space of vacua $(\S^1)^5/\Z_2$, then the quantization of the
bosons $\alpha_i$ gives a unique ground state $\Psi$ with
wavefunction $\Psi(\alpha_i)=1$.  Degeneracy
must come from fermion zero modes.  The hypermultiplet has four
fermion helicity states (coming from quantization of eight fermi fields),
each of which has a zero mode that can be filled or empty.  This
gives $2^4=16$ states which can easily be seen to have the quantum numbers
of a single $N=4$ vector multiplet.  The $U(1)$ vector multiplet
has again four fermion helicity states that give zero modes
which for a given value of the $\alpha_i$ can be either filled or empty.
It may appear therefore that we are headed for a sixteen-fold degeneracy
again.  However,
these  zero modes  are all odd under
$w$ so only the eight
states with even occupation number survive the projection
onto $w$-invariant states (of course, $\Psi$ is $w$-invariant, so the
fermionic part of the wave-function must be invariant also);
these eight states are bosonic and can be seen to transform as a vector
of the transverse $SO(8)$ of light cone gauge.
So ignoring the singularities,
the  quantization gives the eight supermultiplets of type (a)
mentioned above.

The remaining sixteen supermultiplets of type (b) must
 come from the singularities.
Note in particular that the sixteen singularities of type (2) mentioned
above are in natural correspondence with the sixteen unbroken
$U(1)$'s left by the $SO(32) $ Wilson lines, just like the missing
sixteen vector multiplets of type (b), so plausibly
 each singularity of type (2) produces one vector multiplet.
A ``singularity'' of type (2) is just a point in $({\bf S}^1)^5/{\bf Z}_2$
at which a  charged hypermultiplet becomes massless.
To compute the effect of the singularity, one simply includes
the hypermultiplets in the quantization.  As the hypermultiplet
is only relevant very near the ``singularity,'' one can
replace $({\bf S}^1)^5/{\bf Z}_2$ by $\R^5$, which it looks
like locally.
$\R^5$ can be interpreted as the classical space of ground states
of six-dimensional $U(1)$ gauge theory dimensionally reduced
(not compactified) to one dimension (time), so the quantum
mechanics near the ``singularity'' can be interpreted as the
dimensionally reduced $U(1)$ gauge theory with the charged
hypermultiplet.
A supersymmetric state localized near the ``singularity,'' which would
have been missed in the above naive analysis and could give the
missing vector multiplet of type (b), is simply
a normalizable supersymmetric ground state in the $U(1)$ problem;
it would be a ``bound state at threshold,'' since the $U(1)$ problem,
because of non-compactness of $\R^5$ and vanishing potential at
infinity, has a continuum of states starting at zero energy.
Bound states at threshold being notoriously tricky, the analysis
will not be attempted here.  But it is hopefully now clear that
the richer world-volume structure of the five-brane claimed in this
paper is an essential piece of solving the $H$-monopole problem.

I would like to thank J. Polchinski for discussions.

\listrefs
\end